# Puzzle Encryption Algorithm

Gregory Alvarez[1], Charles Berenguer[1]

[1] Goswell Security Laboratory
35 rue Lauriston, 75016
Paris, France
http://www.goswell.net

gregory.alvarez@goswell.net
charles.berenguer@goswell.net

September 2, 2013

**Abstract**

This document describes the symmetric encryption algorithm called Puzzle. It is free and open. The objective of this paper is to get an opinion about its security from the cryptology community. It is separated in two parts, a technical description of the algorithm and its cryptanalysis.

The algorithm has some interesting properties :

- The block size is variable and unknown from an attacker.
- The size of the key has no limit and is unknown from an attacker.
- The key size does not affect the algorithm speed (using a 256 bit key is the same as using a 1024 bit key).
- The algorithm is much faster than the average cryptographic function. Experimental test showed 600 Mo/s - 4 cycles/byte on an Intel Core 2 Duo P8600 2.40GHz and 1,2 Go/s - 2 cycles/byte on an Intel i5-3210M 2.50GHz. Both CPU had only 2 cores.

# 1 Introduction

Puzzle is a block cipher algorithm based on transposition rather than substitution. It uses the following formula to reorder the plaintext :

*FinalPosition = (InitialPosition * KeyBytes + OtherKeyBytes) modulo(BlockSize)*

The idea behind this formula is like cutting every letter of a whole book and shuffling them together. The result is so big and mixed up that it is possible to recreate a completely different book from it. This makes reversing the process very hard without the password.

The algorithm does the following steps in order :

- Key generation
- Block Size calculation
- Plaintext XOR
- Map creation
- Encryption / Decryption

The formulas and equation in this document are presented in a C like syntax. The position of an element starts at 0, not 1.

## 1.1 Key Generation

The key is generated by hashing parts of the password and concatenating them together. Every 3 letter (this value is arbitrary) a hash of all the 3 letters groups before it is generated and added to the key. Using "mypassword" as a password would generate the following key ("+" is a concatenation and "Hash()" is a hash function) :

*Key = Hash("myp") + Hash("mypass") + Hash("mypasswor") + Hash("mypassword")*

Since it will be very easy to brute-force a 3 letters hash like Hash("myp"), a second pass in reverse order is necessary to prevent attacks :

*A = Hash("myp")*
*B = Hash("mypass")*
*C = Hash("mypasswor")*
*D = Hash("mypassword")*

*IntermediateKey = Hash(A + B + C + D) + Hash(B + C + D) + Hash(C + D) + Hash(A + B + D)*

The last hash is a hash of all the parts except the one before the last. Puzzle needs 2 keys, one to XOR the plaintext and one to generate the mapping between initial and final position. The second key is generated using the same method as below except with the password in reverse order. For example if the password is "mypassword" the second key will be generated using "drowssapym".

To isolate each key, a last step is done before using them. The intermediate keys are separated in 4 equal parts (*E, F, G* and *H*) and XOR together (+ is a concatenation and ^ is an XOR) :

*IntermediateKey = E + F + G + H*
*FinalKey = (E ^ G) + (E ^ H) + (F ^ G) + (F ^ H)*

The first final key is used for the XOR and the second final key (with the password in reverse order) is used to create the mapping.

The *FirstFinalKey* is used in this document as an array of bytes.

The *SecondFinalKey* is used in this document as an array of unsigned integer (4 bytes). Each value used from the key can go from 0 to 4 294 967 295. When all the value have been used, the first byte is shifted to the end of the array and the position resets.

## 1.2 Block Size Calculation

The algorithm reorder elements. It is possible to consider a byte as one element but also a bit. However, reordering bits is much slower than bytes due to computers memory characteristics. For speed improvement, it is better to reorder bits only when necessary.

The block size is unknown from an attacker. It is calculated using bytes from the *SecondFinalKey*. The user needs to provide a reference block size :

*K[i] = SecondFinalKey[i]*
*BlockSize = ReferenceBlockSize + (K[0] + K[1] + K[2] + K[3] + K[4] + K[5]) modulo(1/2 \* ReferenceBlockSize)*

The minimum block size allowed is 100 elements for a byte mapping (100 bytes) and 128 elements for a bit mapping (16 bytes). The block size of a bit mapping is always a multiple of 8. If speed is an issue, padding plaintext below 100 bytes should be considered instead of switching to a bit mapping.

## 1.3 Plaintext Xor

The plaintext is xored with the first final key. This step is done for obfuscation in concordance with Shannon's principle [1]: confusion and diffusion. The security of the algorithm resides in the difficulty to recover the initial position of an element in the ciphertext. For example, without obfuscation the plaintext "AAAAA" will be equal to his ciphertext. "i" represent the element position in the plaintext and "BlockNumber" the number of the block :

*XoredPlaintext[i] = Plaintext[i] ^ FirstFinalKey[(i + BlockNumber \* BlockSize) modulo(KeySize)]*

## 1.4 Map Creation

A map is the association of the initial positions to their final positions in the plaintext/ciphertext. It can be represented as an array of *BlockSize* length :

*Map[InitialPosition] = FinalPosition*

Or for the decryption :

*Map[FinalPosition] = InitialPosition*

This document describes two methods for generating the map. The first one is more secure due to its highly non-linear characteristics but slower than the second method. The computation power needed to generate the map grow exponentially making very difficult the calculation of maps with big block size.

The second method, however, is more linear but faster. The computation power needed grow linearly allowing the calculation of maps with big block size like 1 Mo and 10 Mo. With small blocks from 100 to 10 000 bytes, the difference of speed between the two methods is negligible. After 10 000 bytes the first method slow down exponentially to get one hundred times more inefficient than the second.

For blocks under 10 000 bytes, the first method should be used (10 000 included). For bigger blocks, the second method should be used.

With message under 100 bytes, the first method should be used with a bit mapping.

**First Method (Memory Unfolding)**

A set is generated containing all the position from 0 to *(BlockSize - 1)* in order:

*PositionAvailable = [ 0; 1; 2; ... BlockSize - 1 ]*

The final position of an element is then extracted from this set using the following formula :

*I = InitialPosition*
*K = SecondFinalKey*
*KP = KeyPosition*

*IntermediatePosition = (I * K[KP modulo(KeySize)] + K[(KP+1) modulo(KeySize)]) modulo(BlockSize)*
*FinalPosition = PositionAvailable[IntermediatePosition modulo(PositionAvailableSize)]*

The element extracted from *PositionAvailable* is then removed and the size of the set decremented. The key position is incremented by two so that the block *I* and *I + 1* don't share any byte of the key :

*KeyPosition = KeyPosition + 2*

Theses steps are done for every initial position of the plaintext.

**Second Method (Iteration)**

The map is calculated using the following formula :

*I = InitialPosition*
*K = SecondFinalKey*
*KP = KeyPosition*

*FinalPosition = (I * K[KP modulo(KeySize)] + K[(KP+1) modulo(KeySize)]) modulo(BlockSize)*

When the *FinalPosition* is already used by another initial position, it is decremented or incremented until finding a free final position :

*FinalPosition = FinalPosition + 1*

or

*FinalPosition = FinalPosition - 1*

The parity of the second key is used to choose between the direction (incrementing or decrementing) :

*Direction = SecondFinalKey[KP modulo(KeySize)] modulo(2)*

If *Direction* is equal to 1, the *FinalPosition* is incremented, otherwise it is decremented. The key position is incremented by two so that the block *I* and *I + 1* don't share any byte of the key :

*KeyPosition = KeyPosition + 2*

Theses steps are done for every initial position of the plaintext.

## 1.5 Encryption and Decryption

Once the mapping is done, every byte of the plaintext gets assigned a different position in the ciphertext. Encryption and decryption are done by moving around bytes in accordance with the map. To prevent differential attacks, each block of plaintext is shifted before being xored and encrypted. The shifted bytes are added at the end of the plaintext. The map is used to calculate their number :

*NumberOfShiftedBytes = Map[BlockNumber]*

To prevent a byte to be twice in the same position (with a same *Map*), a new *Map* is calculated every *BlockSize* block encrypted by regenerating the keys.

## 1.6 Keys Regeneration

Both keys are regenerated. They are separated into blocks of the algorithm hashes lengths and each block is replaced by the value of its own hash. "i" represent the block number :

*Key[i] = Hash(Key[i])*

## 1.7 Mode of Operation

An Initialization Vector must always be used in any mode of operation to prevent having the same map pattern with two streams of data (file, network connexion...) and potentially being vulnerable to differential attacks.

**Packet Network**

A packet network is usually small. It is possible to fragment a bigger message into multiple smaller packets but this case is not universal, and for example, not applicable to VOIP. For this reasons calculating a secret block size is not useful when encrypting communication over a network because in most situations the packet size will be equal to the block size. Having a secret block size is useful when encrypting data bigger than the block size.

**File**

The best way to encrypt a file is to use an IV and a secret block size. The ideal solution to prevent attacks when writing modification will be to re-encrypt the entire file with a new IV. Unfortunately, this method is hard to implement with on-the-fly encryption and larges files. If the file to be encrypted is smaller than the block size, the whole file is considered as one block.

## 1.8 Initialization Vector

Using the algorithm in Electronic Codebook mode is not a good idea. When puzzle is used with a same block size and keys, all the maps will be identical. To prevent that, it is strongly recommended using an Initialization Vector. Since the key size is unknown, the Initialization Vector needs to be extended to match it.

To do so, the original IV is hashed and discarded. It is not part of the extended IV. The result is then hashed and concatenated to itself :

> *ExtendedIV = Hash(IV)*
> *ExtendedIV = ExtendedIV + Hash(ExtendedIV)*

The concatenation part is repeated as needed to extend the IV to the key length. Once done, the extended IV is xored with the two keys before they can be used by the algorithm.

## 1.9 Drawbacks

Encrypting smaller data than the established limit is dangerous. It will make very easy to brute force the ciphertext by reordering it. Bigger is the data, the better.

It is possible to map bits instead of bytes, making the algorithm more secure and allowing to encrypt much smaller data. However, computers are designed to work in groups of bits, making problematic moving one bit around in memory. An implementation of Puzzle to work on bits add more processor operations, reducing the speed of the algorithm.

The encryption/decryption process needs between 2 and 4 times the block size in memory.

## 2 Cryptanalysis

All the cryptanalysis presented are done on a byte mapping. They work the same way with a bit mapping.

## 2.1 Brute Force

There are two ways of brute-forcing the algorithm: generating every key or reordering the ciphertext.

Today the key size of a symmetric encryption algorithm doesn't go higher than 256 bits. This represents $10^{77}$ possible combination. In comparison, there are $10^{80}$ atoms in the observable universe, which is 13 billion ($10^{10}$) years old [2]. Using 50 supercomputers to break a 256 bit key, given that each of them can calculate $10^{18}$ combinations per second, will take $10^{51}$ years [3]. That's more than a billion times the age of the universe. The puzzle algorithm doesn't have any limit on the key size. Using a 5 character password with a 512 bit hash algorithm for the key generation will create a 1024 bit key. These represent $10^{308}$ possible combinations. Since the key size is unknown, an attacker will have to try them all, exponentially increasing the number of possible combinations.

Reordering the ciphertext is quite harder than brute-forcing the key. The number of possible combinations is the factorial of the block size. For the minimum allowed, 100 bytes, the number of possible combinations is $10^{157}$. This is more than a trillion billion times the number of combination for a 256 bit key. The Puzzle algorithm can go up to 1 Mega byte blocks without any trouble, raising the number of possible combinations for reordering the ciphertext to $10^{5\ 565\ 708}$.

## 2.2 Linear Cryptanalysis

The linear attack is based on the knowledge of an initial position / final position pair [4]. However, the Puzzle algorithm doesn't allow to recover the initial position of an element in the ciphertext.

Considering it was possible, the linearity decays exponentially when the initial position of an element increases in the plaintext. Without shifting, the firsts elements of the plaintext would have a good linear probability with the second method. However, since the plaintext is shifted before being xored and mapped, every element gets associated with the final position of another one. It is impossible with a linear equation to link an initial position with a final position, and extract a byte of the key, due to the non-linearity characteristic of the algorithm: the value of the key doesn't respond to the validity of the linear equation.

Even if it was possible, the key size is unknown and multiple value can be correct due to the modulo operation. In this condition, it is impossible to exploit a key byte found by linear cryptanalysis.

## 2.3 Differential Cryptanalysis

Different mechanisms have been implemented to prevent differential attacks. The block size is assumed known (like with packet network). Multiple pairs of plaintext with one byte difference and their corresponding ciphertext are also considered known. The differential cryptanalysis [4] consists in correlating the difference between the plaintext pairs to the difference between their corresponding ciphertext pairs. This attack is not possible :

- The plaintext is shifted before being xored. This implies that the ciphertexts of a same plaintext encrypted multiple times will be entirely different. This characteristic is obtained by not resetting the position of the first final key (used to xor) between blocks and by shifting the plaintext.

- Considering that it is possible to associate an initial position in the plaintext to its final position in the ciphertext, extracting the key is impossible. The non-linearity of the algorithm prevents from recovering the associated part of the key.

- Even if a relation between an initial position and a final position is found, it will only be valid for the current block and can't be reused. A byte doesn't have the same final position twice with the same map.

## 2.4 Differential-linear Cryptanalysis

This attack combines the differential and linear cryptanalysis technique [5]. The differential attack is used to get the final position of a plaintext byte. Considering the attack possible, the linear cryptanalysis technique permits to create the following equation :

*FinalPosition = ((InitialPosition + (BlockNumber \* A + B) modulo(BlockSize)) \* C + D) modulo(BlockSize)*

There are 4 unknown values (4 bytes each) of the key (representing a non continuous 128 bit block) *A*, *B*, *C* and *D*. The *BlockSize* and *BlockNumber* are considered known by the attacker. The *InitialPosition* and *FinalPosition* are discovered by the differential cryptanalysis. There is in average $10^{20}$ valid combinations when *BlockSize* < *B* and *BlockSize* < *D* for a single byte.

Considering the attacker has the computational power to generate this amount of combination, it will be impossible to differentiate the right key from the invalid ones. With the non-linearity characteristic of the algorithm, there is also a very high probability that the relation between the initial position and the final position can't be represented as an equation. This invalidates all the keys found and prevent linear attacks.

## 2.5 Statistical Analysis

The data in appendixes 1 to 5 have been generated using a random 512 bit key for each encryption and a 10 000 byte block. The graphs on the right represent a group of 100 encryptions and the one on the left a single encryption. The appendixes 1 to 3 are maps. The appendixes 4 and 5 show the nonlinear coefficient of the 2 mapping methods.

The nonlinear coefficient represents the number of final positions in the map that doesn't satisfy the algorithm's formula (*FinalPosition = InitialPosition * KeyByte*...).

The appendix 1 shows the *FinalPosition* distribution without iteration or unfolding method. In other words, it is just the result of the formula :

*I = InitialPosition*
*K = SecondFinalKey*
*KP = KeyPosition*

*FinalPosition = (I * K[KP modulo(KeySize)] + K[(KP+1) modulo(KeySize)]) modulo(BlockSize)*

The appendixes 2 and 3 represent maps using the iteration and unfolding method. The iteration effects are visible on the left graph (appendix 3) creating irregularities at the end of the map. As shown by the graphs on the right (appendixes 1, 2 and 3), the final position is uniformly distributed.

The appendix 4 and 5 show the nonlinear coefficient for the mapping methods. It increases at the end of the map due to the higher probability of having a used space (appendix 4, iteration method). The second method doesn't go further that 60% of nonlinearity. However, the unfolding method (appendix 4) has a 99 % nonlinearity (the first element will always be linear).

# Appendix

Appendix 1 - FinalPosition without iteration or unfolding method

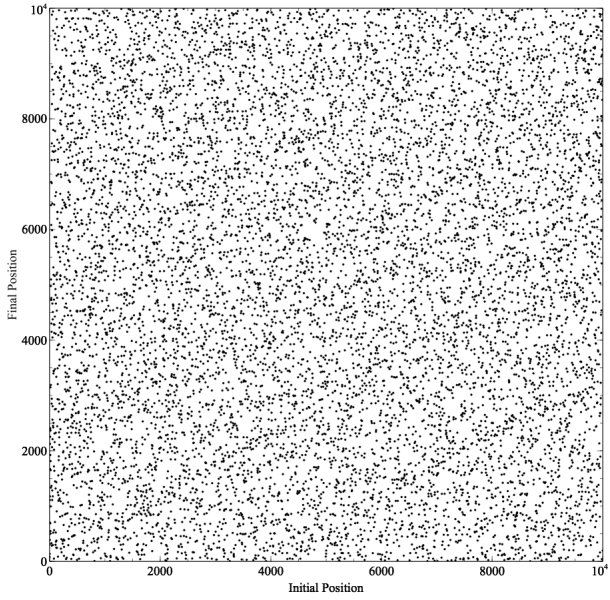 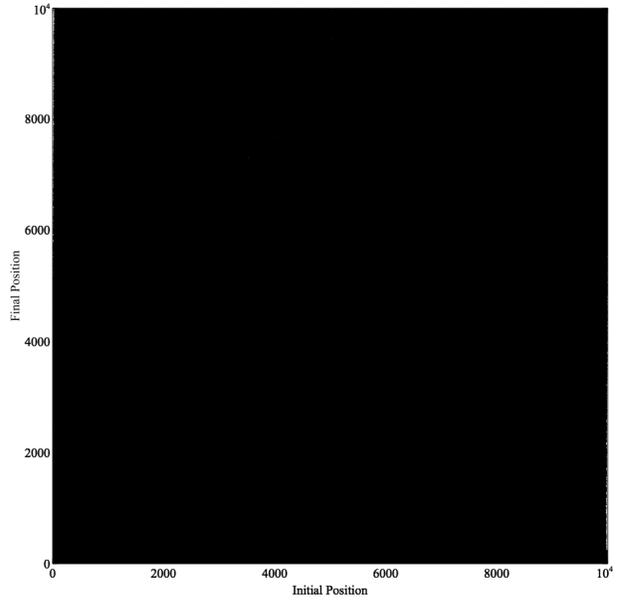

Appendix 2 - FinalPosition with iteration method

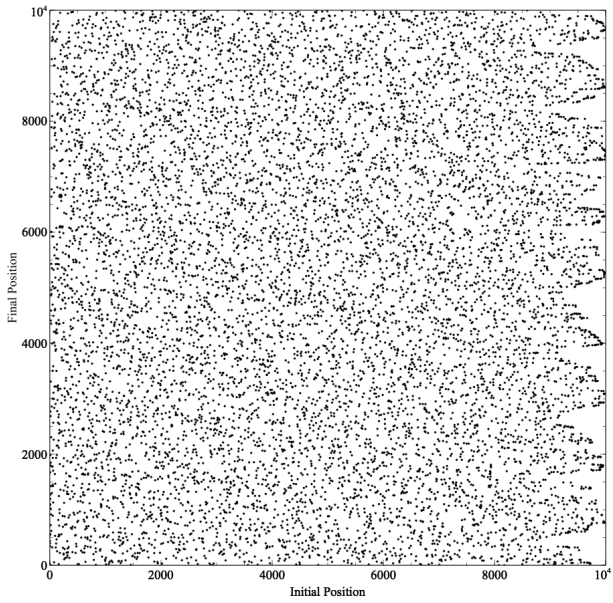 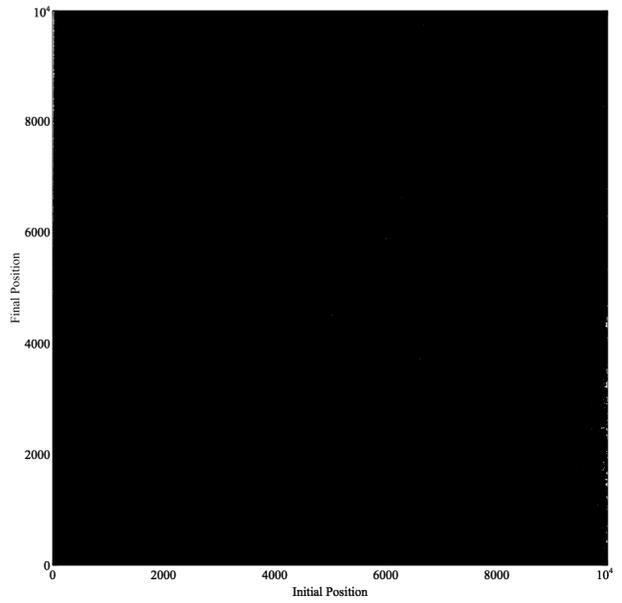

Appendix 3 - FinalPosition with unfolding method

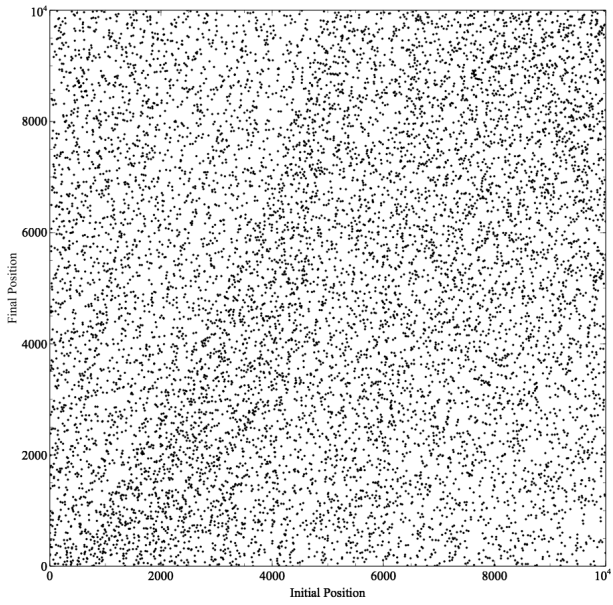
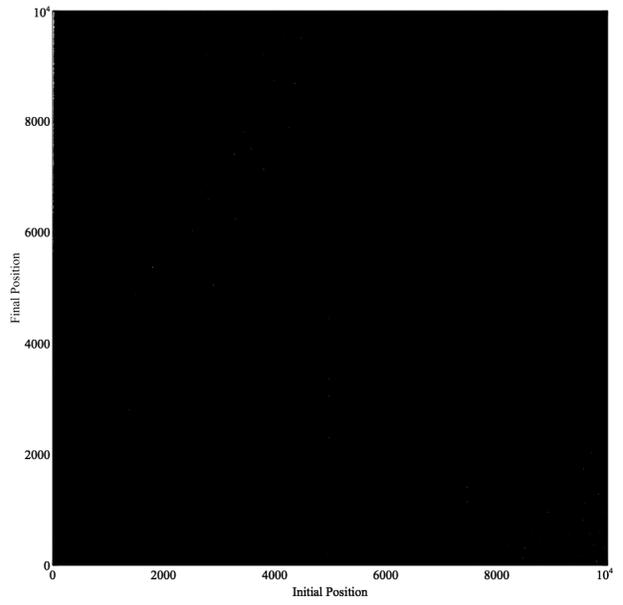

Appendix 4 - Nonlinear coefficient of iteration method

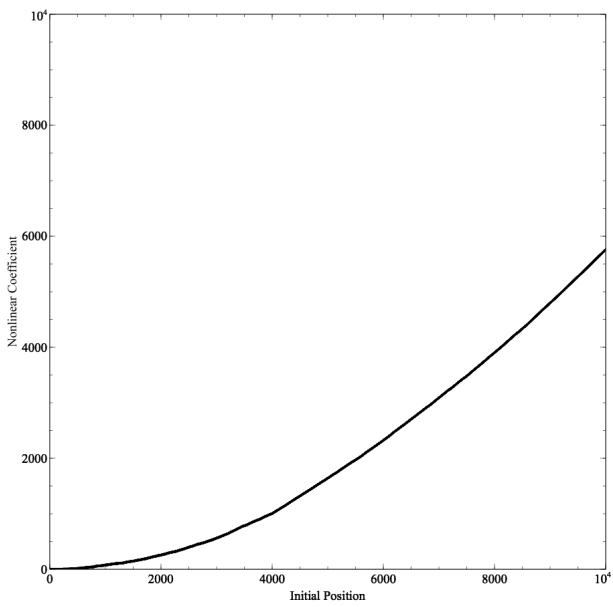
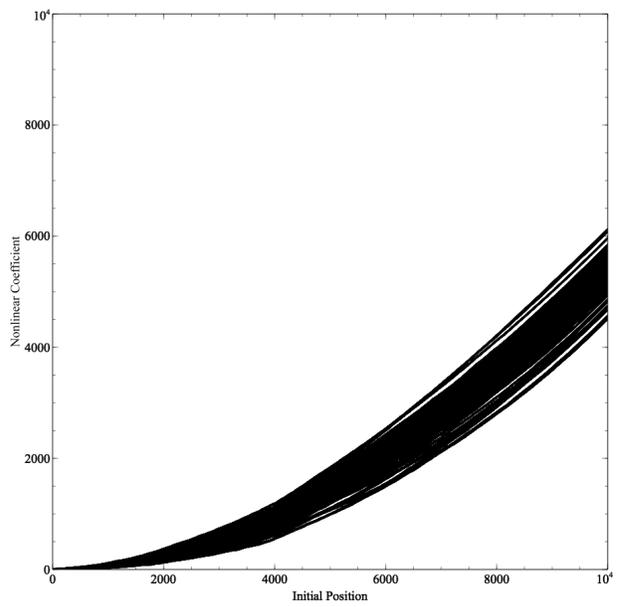

Appendix 5 - Nonlinear coefficient of unfolding method

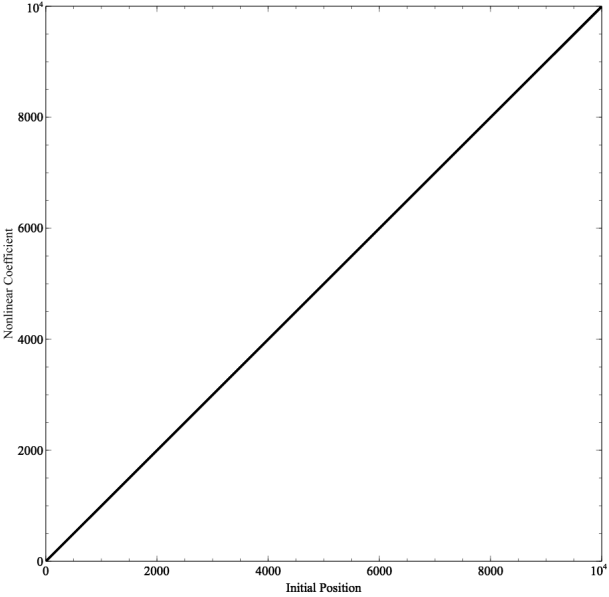 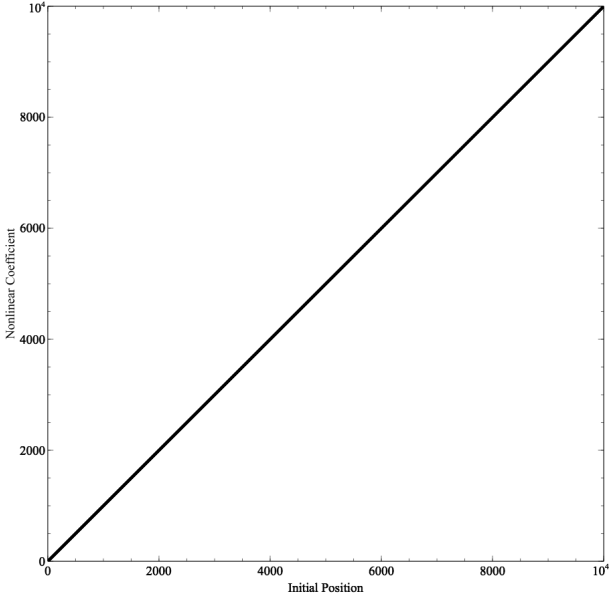